\documentclass[1p]{elsarticle}
\usepackage[utf8]{inputenc}
\usepackage{amsmath}
\usepackage{amsfonts}
\usepackage{mathrsfs}
\usepackage{amssymb}
\usepackage{xfrac}
\usepackage{graphicx}
\usepackage{booktabs}
\usepackage{setspace}

\DeclareMathOperator\arctanh{arctanh}

\makeatletter
\def\ps@pprintTitle{%
 \let\@oddhead\@empty
 \let\@evenhead\@empty
 \def\@oddfoot{}%
 \let\@evenfoot\@oddfoot}
\makeatother

\makeatletter
\def\@author#1{\g@addto@macro\elsauthors{\normalsize%
    \def\baselinestretch{1}%
    \upshape\authorsep#1\unskip\textsuperscript{%
      \ifx\@fnmark\@empty\else\unskip\sep\@fnmark\let\sep=,\fi
      \ifx\@corref\@empty\else\unskip\sep\@corref\let\sep=,\fi
      }%
    \def\authorsep{\unskip,\space}%
    \global\let\@fnmark\@empty
    \global\let\@corref\@empty  
    \global\let\sep\@empty}%
    \@eadauthor={#1}
}
\makeatother

\newcommand{\degrees}{\ensuremath{^\circ}}
\begin{document}

\begin{frontmatter}
\title{{\small A Letter of Intent to Jefferson Lab PAC 44, June 6, 2016}\\Search for Exotic Gluonic States in the Nucleus}

\author{M.~Jones}
\author{C.~Keith}
\author{J.~Maxwell\corref{cor1}}
\ead{jmaxwell@jlab.org}
\author{D.~Meekins}
\address{Thomas Jefferson National Accelerator Facility, Newport News, VA 23606}

\author{W.~Detmold}
\author{R.~Jaffe}
\author{R.~Milner}
\author{P.~Shanahan}
\address{Laboratory for Nuclear Science, MIT, Cambridge, MA 02139}

\author{D.~Crabb}
\author{D.~Day}
\author{D.~Keller}
\author{O.~A.~Rondon}
\address{University of Virginia, Charlottesville, VA 22904}

\author{J.~Pierce}
\address{Oak Ridge National Laboratory, Oak Ridge, TN 37831}

\cortext[cor1]{Corresponding author}

\begin{abstract}
We renew our intent to submit a proposal to perform  a search for a non-zero value of the unmeasured hadronic double helicity flip structure function ${\Delta}(x,Q^2)$, predicted to be sensitive to gluons in the nucleus. This would be performed with an unpolarized electron beam and transversely polarized, spin-1, nuclear target. This structure function was first identified by Jaffe and Manohar in 1989 as ``a clear signature for exotic gluonic components in the target,'' and a recent lattice QCD result by our collaborators has prompted renewed interest in the topic. An inclusive search with deep inelastic scattering, below $x$ of 0.3, via single spin tensor asymmetries may be feasible using the CEBAF 12\,GeV electron beam and JLab/UVa solid polarized target, and would represent the first experimental exploration of this quantity.
\end{abstract}

\end{frontmatter}
\clearpage

\section{Introduction}
Despite the fundamental role the gluon fills in QCD, direct measures of gluonic states in the nucleus remain elusive.  As the gluon does not couple directly to the photon, it is probed only indirectly in electron scattering from hadrons, but the dominance of the gluonic parton distribution function at low $x$ highlights the importance of gluonic interactions in the nucleon. While to first order nuclei are bound states of protons and neutrons, as the spin of the nucleus increases, higher-order behavior in the nucleus becomes available in the form of additional nuclear structure functions.

Jaffe and Manohar \cite{Jaffe} describe a leading twist structure function which is sensitive to gluonic states in the nucleus but free from contributions from the motion and binding of nucleons in the nucleus. 
This quantity, ${\Delta}(x,Q^2)$, is not sensitive to the contributions of bound nucleons or pions in the nucleus, as neither can contribute two units of helicity, and likewise neither can any state with spin less than one contribute. Ma \textit{et al.}\ \cite{Ma} call ${\Delta}(x,Q^2)$, labeled as $\hat{G}_T$ in their paper, the most interesting of the structure functions related to tensor polarization.

\begin{figure}[hb]
\begin{center}
\includegraphics[width=3in]{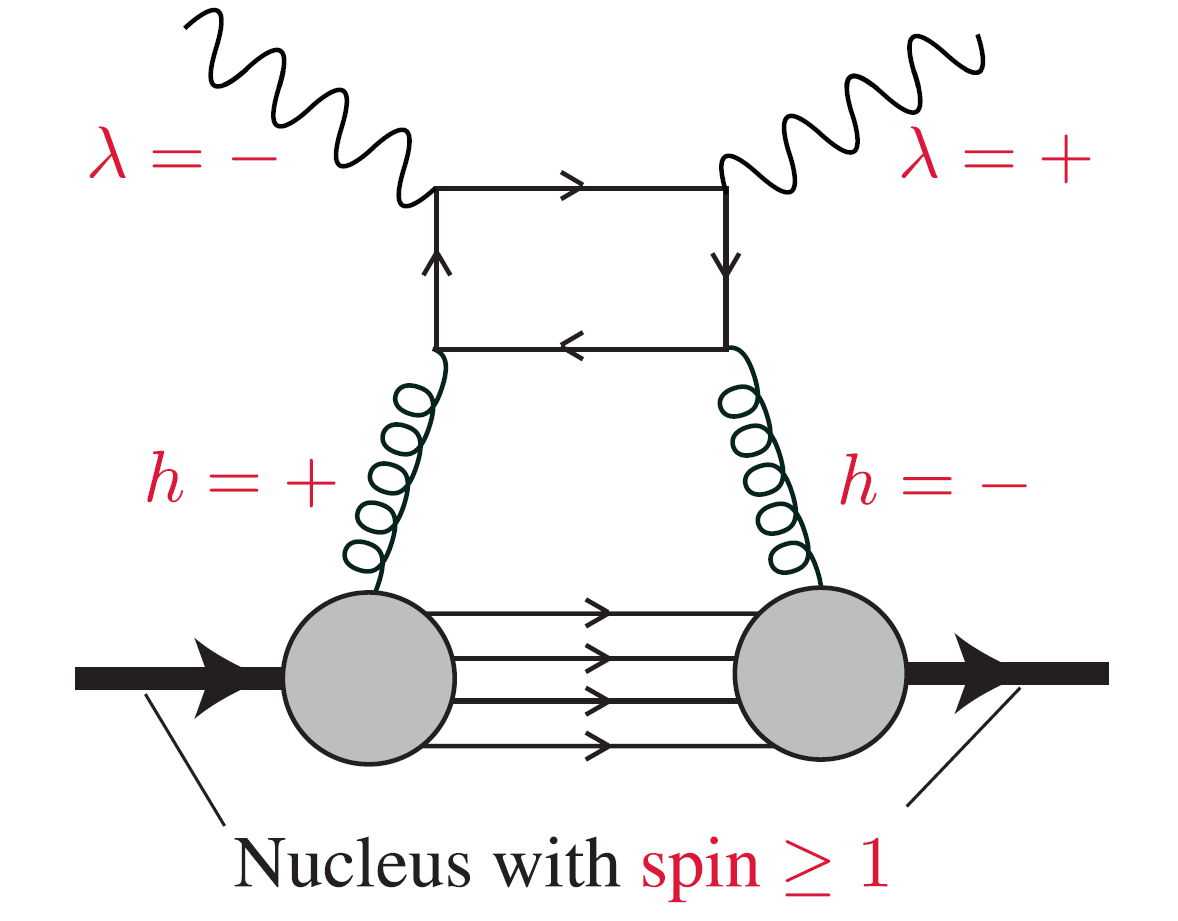}
\end{center}
\caption{Example process to which the photon double helicity flip is sensitive.}
\label{fig:nuglu}
\end{figure}

A measurement of ${\Delta}(x,Q^2)$ at Jefferson Lab would require an unpolarized electron beam incident on a transversely aligned polarized target of spin greater than or equal to 1, and would constitute a pioneering search on a previously unexplored observable.  This measurement would encounter similar experimental challenges to measurements of spin structure function $g_2$, which also requires a transverse target, and tensor structure functions $b_1$, which also requires significant \textit{alignment}, or tensor polarization (see section \ref{sec:tar}). However, such hurdles have been overcome at JLab and SLAC using solid polarized targets such as ammonia and lithium hydride. With no prior measurements to guide a search, we have identified the  gluon dominated low $x$ region as an attractive place to start.

The first challenge given by PAC 42 in response to our 2014 letter was to obtain guidance for our ${\Delta}(x,Q^2)$ measurement from lattice calculations. The search for a non-zero signal for our exotic glue operator, lead by W.~Detmold and P.~Shanahan, bore fruit in April of 2016 \cite{Shanahan}. A clear non-zero signal was found for the first moment of ${\Delta}(x,Q^2)$ on a spin--1 $\phi$ ($s\bar{s}$) meson, here acting as a surrogate for our nuclear target. While these results lack renormalization and other corrections needed to form an estimate for ${\Delta}(x,Q^2)$ in a nucleus, they represent an extremely successful proof of principle demonstrating that this quantity is calculable on the lattice. Essentially no suppression was seen for the double-helicity-flip amplitude relative to the spin-independent amplitude, which bodes well for a measurement. The excitement which these results have generated have led us to reinvigorate our efforts toward a measurement of ${\Delta}(x,Q^2)$ at Jefferson Lab.

\subsection{Theoretical Motivation}

Following reference \cite{Jaffe}, the usual hadronic tensor for inelastic scattering on a spin-1 target 
\begin{equation}
W_{\mu\nu}(E',E)=\frac{1}{4\pi} \int \mathrm{d}^4 x e^{iq \cdot x} \langle p,E' \vert [j_{\mu}(x),j_{\nu}(0)] \vert p,E \rangle
\end{equation}
for polarization vector $E_{\mu}$, can be expressed in terms of a target polarization independent tensor $W_{\mu\nu,\alpha\beta}$:
\begin{equation}
W_{\mu\nu}(E,E')=E'^{*\alpha}E^{\beta}W_{\mu\nu,\alpha\beta}.
\end{equation}
This $W_{\mu\nu,\alpha\beta}$ can in turn be expressed as a sum of helicity projection operators $P(hH,h'H'$)
\ with helicity components of photon and target in the forward axis $h$ and $H = -,0,+$, and the imaginary parts of the corresponding forward Compton helicity amplitude $A_{hH,h'H' }$:
\begin{equation}
W_{\mu\nu,\alpha\beta} = \sum_{hH,h'H'} P(hH,h'H')_{\mu\nu,\alpha\beta}A_{hH,h'H' },
\end{equation} 
with the sum constrained by $h+H=-h'+H'$.

As we are interested in only the double helicity flip component, we can denote $A_{+-,-+}=A_{-+,+-}$ as ${\Delta}(x,Q^2)$, so that the double helicity contribution to $W_{\mu\nu,\alpha\beta}$ is 
\begin{equation}
W_{\mu\nu,\alpha\beta}^{{\Delta}=2} = [P(+-,-+)_{\mu\nu,\alpha\beta}+P(-+,+-)_{\mu\nu,\alpha\beta}]{\Delta}(x,Q^2).
\end{equation}
Expanding the helicity projection operators in terms of the photon and target polarization vectors results in
\begin{align}
W_{\mu\nu,\alpha\beta}^{{\Delta}=2}(E,E') &=  \frac{1}{2}v \left(\left\{ \left[   E^{\prime*}_{\mu}-\frac{q \cdot E^{\prime*}}{\kappa\nu} (  p_{\mu}-\frac{M^2}{\nu} q_{\mu} )\right] \right.\right.\notag\\
&\left.\left.\times\left[E_{\nu}-\frac{q\cdot E}{\kappa\nu}\left( p_{\nu} - \frac{M^2}{\nu}q_{\nu} \right) \right] + (\mu\leftrightarrow\nu) \right\}  \right.\notag\\
&\left.-\left[ g_{\mu\nu}-\frac{q_{\mu}q_{\nu}}{q^2} + \frac{q^2}{\kappa\nu^2} \left( p_{\mu}-\frac{\nu}{q^2}q_{\mu} \right) \left( p_{\nu}-\frac{\nu}{q^2}q_{\nu}\right)\right]\right.\notag\\
&\times \left. \left(E^{\prime*}\cdot E + \frac{M^2}{\kappa\nu^2}q\cdot E^{\prime*} q\cdot E \right) \right){\Delta}(x,Q^2).
\end{align}
This expression will vanish should $E=E'=E^{\pm}_0$ when the target is polarized parallel, or if averaged over spin. 

This expression is much simplified in the Bjorken limit:
\begin{equation}
\label{eq:delta}
\lim_{Q^2\rightarrow \infty} \frac{d\sigma}{dx\,dy\,d\phi} = \frac{e^4ME}{4\pi^2Q^4}\left[xy^2F_1(x,Q^2)+(1-y)F_2(x,Q^2) - \frac{x(1-y)}{2}{\Delta}(x,Q^2)\cos2\phi \right],
\end{equation}
with higher twist terms and vanishing kinematic corrections ignored. Here $\phi$ is the angle between the scattering plane and the target spin orientation. If the target is polarized in the opposite direction, the same cross section is obtained, so that the effect is not sensitive to the polarization of the target, but rather the alignment.


A partonic interpretation of $\Delta(x,Q^2)$ can be defined for a target in the infinite momentum frame with its spin in the $\hat{x}$ direction, perpendicular to momentum. For the probability of finding a gluon with momentum fraction $x$ and linearly polarized in the $\hat{x}$,$\hat{y}$ direction $g_{\hat{x},\hat{y}}(x,Q^2)$, we have 
\begin{equation}
\Delta(x,Q^2) = \frac{\alpha_S(Q^2)}{2\pi}\mathrm{Tr} \mathscr{Q^2}x^2 \int^1_x\frac{dy}{y^3}\left( g_{\hat{x}}(x,Q^2) - g_{\hat{y}}(x,Q^2) \right)
\end{equation}
for quark charge matrix $\mathscr{Q} = \mathrm{diag}(\sfrac{2}{3},-\sfrac{1}{3},-\sfrac{1}{3})$.

Sather and Schmidt \cite{Sather} outline the scaling behavior of $\Delta(x,Q^2)$, and calculate the size of its first moment in the bag model for the spin-\sfrac{3}{2} particle, $\Delta^{++}$:
\begin{equation}
\int_0^1 dx\,x\Delta(x,Q^2) = -0.012\alpha_s(Q^2).
\end{equation}
$\Delta(x,Q^2)$ may prove to be even smaller for a spin-1, nuclear target. Further lattice QCD exploration of heavy mesons should shed light on the moments of $\Delta$ we might encounter in light nuclear targets.

\section{Experiment}
Our investigation into the prospects of a measurement of $\Delta(x,Q^2)$ remains preliminary, however several key requirements have already introduced challenging experimental constraints, particularly in the choice of target. The need for a transversely polarized target brings complications, but successful experiments at JLab and SLAC show that the JLab/UVa solid polarized target presents a dependable solution \cite{Pierce}. The fact that we search for truly nuclear effects leads us 
toward heavier nuclei. Nitrogen and lithium offer promising target candidates, as they may be polarized in commonly used $^{14}$NH$_3$ and $^6$LiH, albeit at lower absolute polarization compared to the protons themselves.


\subsection{Method}
For a spin--1 target polarized at angle $\theta_m$ from the $z$-axis and electron incident from $-z$, we can express the differential cross section for the target spin in the $\hat{m}$ direction $\lambda_m = (1,0,-1)$ as:
\begin{align}
\frac{d\sigma}{dx\,dy\,d\phi}(\lambda_m) &= \frac{2 y \alpha^2}{Q^2} \left( F_1+ \frac{2}{3}a_m b_1 + \frac{1-y}{xy^2} \left(F_2+\frac{2}{3}a_m b_2 \right) \right.\notag\\
&\left.- \frac{1-y}{y^2} b_m \sin^2 \theta_m \Delta(x,Q^2) \cos(2\phi) \right)
\end{align}
where
\begin{align*}
a_m &= \frac{1}{4}b_m(3\cos^2\theta_m -1)\\
b_m &= 3|\lambda_m| -2
\end{align*}
for scattered electron angle $\theta$, azimuthal angle $\phi$ with respect to the place of the incident electron and target spin, and usual tensor structure functions $b_1$ and $b_2$ \cite{JaffeEmail}.

If we average over polarization: $N_+ + N_-+N_0 \Rightarrow \bar{\sigma}$, where $b_+ + b_- + b_0 = 0$ we see no dependence on $\Delta$:
$$
\frac{d\bar{\sigma}}{dx\,dy\,d\phi} = \frac{2y\alpha^2}{Q^2} \left( F_1 + \frac{1-y}{xy^2}F_2\right).
$$ Since the differential cross section depends only on the absolute value of the target helicity 
$\lambda_m$, $\Delta$ also cancels out of the usual vector polarization difference $(N_+ - N_0)+(N_0-N_-) = N_+-N_-$.

If we instead take the portion sensitive to \textit{tensor polarization}: $(N_+ - N_0)-(N_0-N_-) = N_+ + N_- - 2N_0 \Rightarrow \Delta\sigma$, where $b_+ + b_- - 2b_0 = 6$, we have 

\begin{align}
\frac{d\Delta\sigma}{dx\,dy\,d\phi} &= \frac{2y\alpha^2}{Q^2} \left( (3\cos^2\theta_m - 1 ) (b_1 + \frac{1-y}{xy^2}b_2) \right. \notag\\
&-\left. 6 \frac{1-y}{y^2}\sin^2\theta_m \Delta(x,Q^2)\cos(2\phi)\right). \label{eq:tensor}
\end{align}

This presents us three ways to measure $\Delta(x,Q^2)$:

\begin{enumerate}
\item Following equation \ref{eq:tensor}, we can take a tensor asymetry measurement of the form\begin{equation}
\mathcal{A} = \frac{1}{A}\frac{N_+ + N_- - 2N_0}{N_+ + N_- + 2N_0}
\end{equation}
which would be made from yields with target polarized in the $m=-1,0,1$ substates $N_-$, $N_0$, and $N_+$ and tensor alignment $A$. Here measurements at each target helicity state are separated in time, making the measurement sensitive to systematic drifts in detectors efficiencies, luminosity, etc.\ over time.

\item We can similarly form a difference between polarized and unpolarized cross sections $$N_+ - \bar{N} = N_+ - \frac{1}{3}(N_+ +N_-+N_0) = \frac{1}{3}(N_+ - N_0) \Rightarrow \hat{\sigma}$$
giving a similar expression to equation \ref{eq:tensor}:
\begin{align}
\frac{d\hat{\sigma}}{dx\,dy\,d\phi} &= \frac{2y\alpha^2}{Q^2} \left( \frac{1}{6} (3\cos^2\theta_m - 1 ) (b_1 + \frac{1-y}{xy^2}b_2) \right. \notag\\
&-\left.  \frac{1-y}{y^2}\sin^2\theta_m \Delta(x,Q^2)\cos(2\phi)\right). \label{eq:diff}
\end{align} 
While $\Delta(x,Q^2)$ is not sensitive to the difference $N_+ - N_-$ of vector polarized states, this difference of polarized and unpolarized cross sections gives us a back door of sorts, allowing a measure of a tensor observable without needing a highly tensor aligned target. The relative ease of producing vector polarization makes this option seem attractive, although the added difficulty of a cross section measurement compared to an asymmetry will force careful consideration and study of options 1 and 2 in simulation.

\item We can leverage the $\cos(2\phi)$ term to produce an azimuthal asymmetry. Fixing the target polarization and measuring the coefficient of $\cos(2\phi)$ would allow a measurement of $\Delta(x,Q^2)$, however, this is difficult with a fixed target at JLab. Standard equipment in Halls A and C offer little to no out of plane acceptance, and a transversely polarized target that will take significant electron beam current does not exist in Hall B. The ability to rotate the JLab/UVa target vertically would warrant reassessment of our method, as would the advent of detector systems which give greater acceptance in $\phi$ in Halls A or C. Approved experiment E12-11-108 uses the transversely polarized JLab/UVa solid target with the SoLID spectrometer in Hall A, so this may offer such a path.
\end{enumerate}

For methods 1 and 2, we see from equations \ref{eq:tensor} and \ref{eq:diff} that $\Delta(x,Q^2)$ would be overwhelmed by the presumably large contributions from tensor structure functions $b_1$ and $b_2$, unless we choose the target helicity angle such that $(3\cos^2\theta_m - 1 ) = 0$, where $\theta_m = 54.7\degrees$. The "Hall B" magnet available for solid polarized target experiments at JLab, last used for a measurement of $g_2^p$ in Hall A in 2012, has 50\degrees \ to 73\degrees \ blocked by superconducting coils, so a new magnet arrangement would be necessary to reach 54.7\degrees.

We are pursuing the possibility of a measurement of ${\Delta}(x,Q^2)$ at Jefferson Lab using the UVa solid polarized target in Hall C following methods 1 and 2 above. The electron polarization from the CEBAF beam would be averaged over, as both approaches require the switching of the target polarization over the course of the experiment. Careful control of systematic drifts over long periods of time would thus be necessary, much as is required in the $b_1$ experiment, E12-13-011, which is conditionally approved for the 12\,GeV running period. Should this measurement mature into an approved experiment, running contiguously with E12-13-011 while the polarized target is installed in Hall C could be convenient.


As discussed in the section \ref{sec:tar}, nitrogen-14 in ammonia is currently the most attractive target material. Should sufficient tensor polarization on nitrogen be achieved in the course of our target development, the tensor structure function $b_1$ on nitrogen could offer an exciting auxiliary measurement within the scope of this effort. As the JLab/UVa target can be rotated in minutes, switching from transverse target polarization to measure $\Delta$, to parallel target polarization to measure $b_1^N$ would be a simple matter.


We have undertaken a cursory study of the kinematics of the measurement, simulating the transversely polarized solid ammonia target scattering into Hall C's SHMS at 10.5 and 15\degrees. Although we show some example kinematics and rates in table \ref{tab:kine}, these do not reflect the 54.7\degrees\ target angle needed to cancel out $b_1$ and $b_2$, and so are here given only for guidance.

\begin{table}
\begin{center}
\begin{tabular}{cccccc}
\toprule
$\theta$ & E (GeV) & E' (GeV) &  $Q^2$ (GeV/c$^2$) & $x$ & Rate (Hz)\\
\midrule
10.5 & 11 & 5 & 1.842 & 0.164 & 170  \\
10.5 & 11 & 4 & 1.474 & 0.112 & 152 \\
10.5 & 11 & 3 & 1.105 & 0.074 & 138 \\
10.5 & 11 & 2 & 0.737 & 0.044 & 100 \\
\midrule
15 & 11 & 5 & 3.748 & 0.333 & 28  \\
15 & 11 & 4 & 2.999 & 0.228 & 30 \\
15 & 11 & 3 & 2.249 & 0.15  & 32  \\
15 & 11 & 2 & 1.499 & 0.089 & 34 \\
\bottomrule
\end{tabular}
\end{center}
\caption{Preliminary table of rates for given HMS Monte Carlo settings for the transversely polarized target in Hall C.}
\label{tab:kine}
\end{table}

\subsection{Transversely Polarized Target}
\label{sec:tar}
We require a target material with a high density of spin--1 nuclei that can be polarized at an angle to the incident beam, and can stay polarized under significant beam current over time. The workhorse materials of JLab's solid polarized target program, ammonia ($^{14}$NH$_3$) and lithium hydride ($^6$LiH), offer robust performance at high degrees of proton polarization in CEBAF's electron beam. The spin--1 $^{14}$N and $^6$Li nuclei in these materials may be the polarized target species we need, with some further development.

Dynamic nuclear polarization (DNP) \cite{Crabb} leverages the high electron polarization in a material at low temperature (1\,K) in a high magnetic field (5\,T) to pump polarizing transitions in the nuclei of interest. In a material with sufficient unpaired electron spins, applying a microwave field at specific frequencies can induce flip-flop transitions of coupled the electron and nuclear spins. The nuclear polarization diffuses through the material, and the unpaired electron spin relaxes, becoming available for a flip-flop with another nucleus. In practice, the unpaired electrons are provided via paramagnetic radicals throughout the target media, which in the case of ammonia \cite{Meyer} can be created via irradiation doping.

For a spin 1 target, the three available magnetic sublevels $I_z = -1,0,+1$ with populations $n_-$, $n_0$ and $n_+$, give rise to both vector \textit{polarization} $P= (n_+-n_0)+(n_0-n_-) = n_+ - n_-$ and tensor \textit{alignment} $A=(n_+-n_0)-(n_0-n_-)=1-3n_0$. When the spin is in thermal equilibrium with the lattice, the alignment can be determined directly from the polarization, neglecting a small quadrupole term:
\begin{equation}
A = 2-\sqrt{4-3P^2}.
\end{equation}

\begin{figure}
\begin{center}
\includegraphics[width=3.5in]{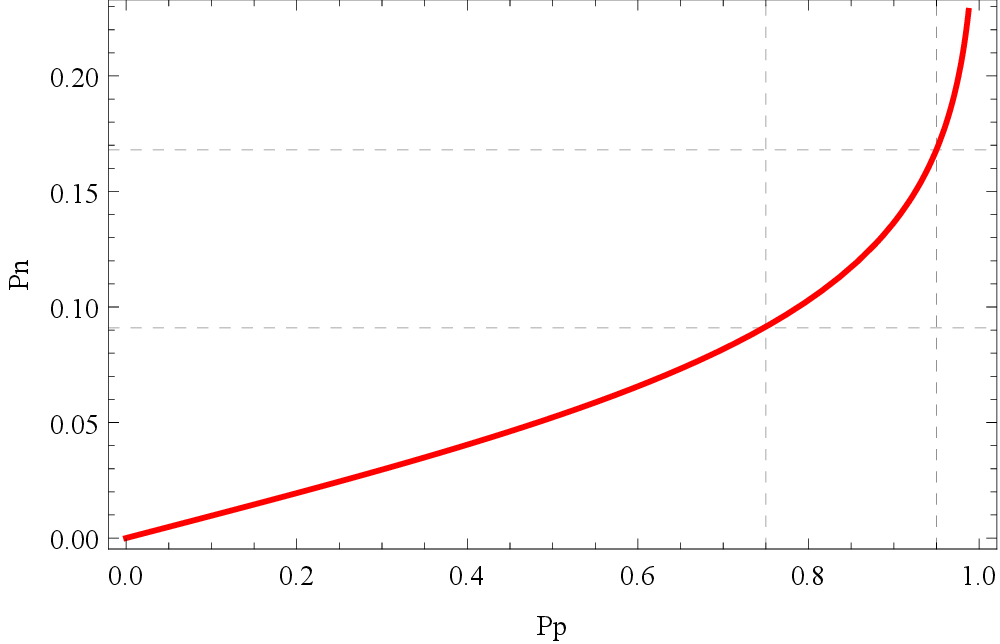}
\end{center}
\caption{Nitrogen polarization as a function of proton polarization in $^{14}$NH$_3$, under DNP at equal spin temperature.}
\label{fig:nit}
\end{figure}

In $^{14}$NH$_3$, the polarization of the  spin-1/2 hydrogen $P_p$ and spin-1 nitrogen $P_N$ are related as 
\begin{equation}
P_N = \frac{4\tanh((\omega_N/\omega_p)\arctanh(P_p))}{3+\tanh^2((\omega_N/\omega_p)\arctanh(P_p))}
\end{equation}
when the species have a common spin temperature, for Larmor frequency $\omega$. Figure \ref{fig:nit} shows this relation, and highlights the difficulty of polarizing via equal spin temperature. Even achieving excellent proton polarization at 95\% will give only 17\% vector polarization in nitrogen, and just 2\% tensor alignment.  

Several techniques can be used to enhance tensor polarization over that available at equal spin temperature. Adiabatic fast passage \cite{Hautle}, and RF hole burning are two such techniques, but in general these would work best with a frozen spin target rather than one which is dynamically polarized. A system such as Hall B's FROST target \cite{Keith} would be an alternative in this case, but it can not maintain polarization in an electron beam. 

Directly improving the vector polarization will thus be crucial for this measurement, whether it is performed via tensor asymmetry, or difference in vector polarized and unpolarized cross sections. Key techniques to investigate are RF spin transfer from another species in the target \cite{Delheij} and spin cross-relaxation by varying the magnetic holding field, which both leverage the same mechanism. Significant polarization in nitrogen has been achieved by the SMC collaboration using cross-relaxation to take the ammonia system out of equal spin temperature \cite{Adeva}. Nitrogen polarization of 40\% was reached by crossing resonances with magnetic field variation. At equal spin temperature, this relatively high polarization would translate to 12\% alignment. However, this technique is likely to benefit from further development, particularly in the 5\,T field of the JLab/UVa target which is twice that used by SMC.

Another complication is the measurement of the nitrogen polarization. The degree of polarization and alignment of a population of spin-1 nuclei can be measured through the two absorption peaks of nitrogen which result from quadrupole coupling by taking a ratio of the peak heights. At 2.5\,T, the Larmor frequency of nitrogen nuclei is 7.7\,MHz, and the quadrupole coupling constant is 0.4\,MHz in ammonia at low temperature. The absorption peaks are then 2.4\,MHz apart and not accessible by one frequency sweep of an NMR Q-meter. This has been overcome by the SMC collaboration using separate sweeps to measure each NMR peak and Q-curves by altering the magnetic field. To our knowledge this has not been done at 5\,T, and would be very disruptive to experimental running. We are pursuing remote tuning NMR techniques to facilitate the rapid measurement of the nitrogen polarization.

The target angle constraint of 54.7\degrees\ which would cancel the contributions of $b_1$ and $b_2$ is an inconvenience, but not out of reach. While the existing polarized target magnet has 54.7\degrees\ blocked by coils, an opening angle to 70\degrees\ on a Helmholtz pair magnet should be possible \cite{Desportes} should a new magnet be built. Designing and sourcing a new superconducting 5\,T magnet is a high priority task in pursuing this measurement.

Ammonia appears to offer an experimentally tenable target, although further development is warranted. RF spin transfer could offer 40\% polarization and 12\% alignment, which may already be sufficient for an exploratory measurement. While ammonia has been focused upon thus far due to its long and reliable history with solid targets, we are pursuing other candidate materials for this unique challenge.


\section{Path Forward}

The hadronic double helicity flip structure function could offer a tantalizing glimpse into gluon states in the nucleus which is inaccessible by other methods. Significant challenges face such a measurement, particularly in the small expected size of the quantity, and in finding the combination of a transversely polarized target of spin-1 nuclei that has significant polarization and tensor alignment.  As it stands, polarized nitrogen in $^{14}$NH$_3$ using the 12\,GeV CEBAF beam, observed with the Hall C SHMS and HMS appears feasible, particularly should there be improvements in the achievable nitrogen polarization. The rates expected in the kinematic region of interest mean sheer statistics could be used to overcome poor alignment. However, this could be plagued with systematic effects from forming an asymmetry with yields at different polarization settings separated by days. 

The discovery potential of a measurement of ${\Delta}(x,Q^2)$ begs continued development, which we intend to pursue. Our collaboration will include some of the world's experts on solid polarized targets, and we will start our effort with a search for a material or technique that may improve the nitrogen polarization or tensor alignment. The exciting lattice calculation results should lead to guidance for the size of  ${\Delta}(x,Q^2)$ in a nucleus-like system. Pursuing these options, as well as searching for alternative methods of measurement to reduce systematic error, should lead us to the submission of an experimental proposal for a future PAC.

\section*{References}
\bibliographystyle{elsarticle-num}
\bibliography{update_draft_LOI}

\end{document}